# Exact solution for a periodically driven magnetic multilayer system

P. I. Naumkin[1], A.V. Nikolaev[2], L.L. Tao[3], M. Ye. Zhuravlev[4]

[1]Centro de Ciencias Matemáticas, UNAM Campus Morelia, AP 61-3 (Xangari), CP 58089 Morelia, Michoacán, México

[2]Skobeltsyn Institute of Nuclear Physics, Moscow State University, Moscow 101000, Russia

[3]School of Physics, Harbin Institute of Technology, Harbin 150001, China

[4]St. Petersburg State University, St. Petersburg 199034, Russia

**Abstract**.

Periodic driving serves as an effective method for controlling the properties of physical systems. Called ‘Floquet engineering,’ it is a broad field of theoretical and experimental activity. Whereas original Floquet theory was proposed to a system of ordinary differential equations, the quantum systems with time-dependent potential require using partial differential equations. Among different methods of analysis of such systems, time series is a most common one. Though general scheme was developed in a number of works, its application to specific problems often faces significant difficulties. In particular, the class of the problems describing magnetic multilayers with time-dependent potential (e.g., rotating magnetization of some of the layers) leads to significant complication of the problem due to two-component wave function and matching conditions at the interfaces. Taking as an example a two-layer system containing magnetic layer with rotating magnetization, we construct a class of solution containing arbitrary but finite number of the terms. The structure of the solution is analyzed. In particular, we show that boundary conditions, which seem a natural generalization of that for a stationary problem, cannot be imposed in the case of rotating magnetizations.

## I. Introduction

Though an original theory of the systems of linear ordinary differential equations with periodic coefficients was developed by Gaston Floquet in 1883 [1], various applied problems still require special analysis of different modifications of the equations with periodic coefficients. For example, periodic driving serves as an effective method of controlling properties of physical systems. Called “Floquet engineering” [2], it is a broad field of theoretical and experimental activity. In particular, oscillating perturbations are used in many experiments with multilayered systems. Such systems include multilayers with magnetic [3,4] and superconducting [5,6] layers. Along with experimental research in this field, the theory of such systems was developed in a number of works. Quantum systems described by time-dependent Schrodinger equation which is a partial differential equation invites the application of the theory of partial differential equations. For the latter case, Floquet-Schrodinger formalism was developed in [7,8] and subsequent works (e.g., [9,10]). The class of the problems describing magnetic multilayers with time-dependent potential (e.g., rotating magnetization of some of the layers) appears to be seriously complicated due to two-component wave function and matching conditions at the interfaces. A significant number of proposed approximate methods [9, 11, 12] confirms the complexity of the problem.

In the present article, we construct the solution to a system of two partial differential equations. These are non-equilibrium Schrodinger equations for a two-layer system consisting of a magnetic layer with rotating magnetization and a non-magnetic layer with spin-orbit coupling. One of the most common forms of the solution to Schrodinger equation with periodic potential is to represent the solution as a series of $\exp(in\omega t)$, where $\omega = 2\pi / T$, where $T$ is the period of the potential. We use a similar representation to construct a class of exact solutions of the problem. The solutions we propose contain finite number of time exponents. We analyze the structure of the obtained solution.

## II. The origin of the problem

The equations which are analyzed in this work arise from the problem of solution of Schrodinger equation for a two-layer system. The system contains a ferromagnetic layer with a rotating magnetization with frequency $\omega$ and a layer with Dresselhaus and bulk Rashba spin-orbits couplings. The initial Schrodinger equations are following:

$$
\begin{aligned}
& i\hbar\partial_t \Psi(x,y,z,t) = -\frac{\hbar^2}{2m}\begin{pmatrix} \Delta & 0 \\ 0 & \Delta \end{pmatrix}\Psi + V\Psi, \\
& V = \begin{pmatrix} J_1 \cos\omega t & -i(J_0/2) + J_1 \sin\omega t \\ i(J_0/2) + J_1 \sin\omega t & -J_1 \cos\omega t \end{pmatrix}, z < 0, \\
& V = \begin{pmatrix} -U_0 & -\alpha\partial_x - i\beta\partial_y \\ \alpha\partial_x - i\beta\partial_y & -U_0 \end{pmatrix}, 0 \le z \le b,
\end{aligned}
\tag{1}
$$

where $J_0, J_1$ are exchange parameters, $\alpha, \beta$ are spin-orbit coupling parameters, and $U_0$ is the difference between the bands' positions in the layers. Choosing

$$
\Psi(x,y,z,t) = e^{i(k_x x + k_y y)} \cdot \begin{pmatrix} \Psi^{\uparrow}(z,t) \\ \Psi^{\downarrow}(z,t) \end{pmatrix}
\tag{2}
$$

and applying transform

$$
H \to (R_0)^{-1} H R_0 \text{ with } R_0 = \frac{1}{2}\begin{pmatrix} 1+i & 1+i \\ -1+i & 1-i \end{pmatrix},
\tag{3}
$$

we get the system of equations in $z$ - and $t$ - variables. Renormalizing the parameters to get dimensionless variables

$$t=\hbar J_0^{-1}\tilde{t},\quad \partial_t=(\partial_t\tilde{t})\partial_{\tilde{t}}=\hbar^{-1}J_0\partial_{\tilde{t}},\quad z=\frac{\hbar}{\sqrt{2mJ_0}}\tilde{z},\ \partial_z^2=(\partial_z\tilde{z})^2\partial_{\tilde{z}}^2=\left(\sqrt{\frac{2mJ_0}{\hbar^2}}\right)^2\partial_{\tilde{z}}^2,$$
$$J_1J_0^{-1}\to J_1,\ U_0J_0^{-1}\to U_0,\ \varepsilon_{xy}J_0^{-1}\to\varepsilon_{xy},\ \alpha k_xJ_0^{-1}\to\alpha k_x,\ \beta k_yJ_0^{-1}\to\beta k_y,\ \tilde{\omega}=\hbar J_0^{-1}\omega \quad (4)$$

and returning to the initial notation, we obtain the following systems of equations:

$z\le 0$

$$\begin{cases}-i\partial_t\Psi^{\uparrow}-\frac{1}{2}\partial_z^2\Psi^{\uparrow}+(\varepsilon_{xy}+\frac{1}{2})\Psi^{\uparrow}+J_1e^{-i\omega t}\Psi^{\downarrow}=0\\ -i\partial_t\Psi^{\downarrow}-\frac{1}{2}\partial_z^2\Psi^{\downarrow}+(\varepsilon_{xy}-\frac{1}{2})\Psi^{\downarrow}+J_1e^{i\omega t}\Psi^{\uparrow}=0\end{cases} \quad (5)$$

$0\le z\le b$

$$\begin{cases}-i\partial_t\Phi^{\uparrow}-\frac{1}{2}\partial_z^2\Phi^{\uparrow}+(\varepsilon_{xy}-U_0+\alpha k_x)\Phi^{\uparrow}-i\beta k_y\Phi^{\downarrow}=0\\ -i\partial_t\Phi^{\downarrow}-\frac{1}{2}\partial_z^2\Phi^{\downarrow}+(\varepsilon_{xy}-U_0-\alpha k_x)\Phi^{\downarrow}+i\beta k_y\Phi^{\uparrow}=0\end{cases} \quad (6)$$

We intend to solve the systems of the equations (5), (6).

## III. Construction of the solution

Let's set $\Psi^{\uparrow}(z,t)=u_1(z,t)e^{-\frac{1}{2}i\omega t-i\Omega t}$, $\Psi^{\downarrow}(z,t)=u_2(z,t)e^{\frac{1}{2}i\omega t-i\Omega t}$, and search the solution in the form $u_1(z,t)=v_1(t)e^{iqz}$, $u_2(z,t)=v_2(t)e^{iqz}$. So, we get

$$\begin{aligned}&-i\partial_t v_1+(\varepsilon_{xy}-\Omega+\tfrac{q^2+1-\omega}{2})v_1+J_1v_2=0\\ &-i\partial_t v_2+(\varepsilon_{xy}-\Omega+\tfrac{q^2-1+\omega}{2})v_2+J_1v_1=0\end{aligned} \quad (7)$$

Substituting

$$\begin{pmatrix}v_1(t)\\ v_2(t)\end{pmatrix}=\begin{pmatrix}v_1\\ v_2\end{pmatrix}e^{-i\lambda t}, \quad (8)$$

we get

$$\begin{vmatrix}\varepsilon_{xy}-\Omega+\frac{q^2+1-\omega}{2}-\lambda & J_1\\ J_1 & \varepsilon_{xy}-\Omega+\frac{q^2-1+\omega}{2}-\lambda\end{vmatrix}=0 \quad (9)$$

$$\begin{aligned}&\lambda^2-\lambda(2\varepsilon_{xy}-2\Omega+q^2)+(\varepsilon_{xy}-\Omega+\tfrac{q^2}{2})^2-\tfrac{(1-\omega)^2}{4}-J_1^2=0,\\ &\lambda_{\pm}=\varepsilon_{xy}-\Omega+\tfrac{q^2}{2}\pm\frac{1}{2}\sqrt{(1-\omega)^2+4J_1^2}\end{aligned} \quad (10)$$

Therefore,

$$\begin{pmatrix} v_1^+ \\ v_2^+ \end{pmatrix} = \begin{pmatrix} K_+ \\ 1 \end{pmatrix}, K_+ = -\frac{2J_1}{1-\omega+\sqrt{(1-\omega)^2+4J_1^2}},$$
$$\begin{pmatrix} v_1^- \\ v_2^- \end{pmatrix} = \begin{pmatrix} K_- \\ 1 \end{pmatrix}, K_- = -\frac{2J_1}{1-\omega-\sqrt{(1-\omega)^2+4J_1^2}} \tag{11}$$

Finally, we chose the solution as follows,

$$\begin{pmatrix} v_1(t) \\ v_2(t) \end{pmatrix} = \begin{pmatrix} v_1^+ \\ v_2^+ \end{pmatrix} e^{-i\lambda_+ t} + \begin{pmatrix} v_1^- \\ v_2^- \end{pmatrix} e^{-i\lambda_- t},$$
$$\begin{pmatrix} \Psi^\uparrow(z,t) \\ \Psi^\downarrow(z,t) \end{pmatrix} = \begin{pmatrix} v_1^+ e^{-\frac{1}{2}i\omega t - i\Omega t} \\ v_2^+ e^{\frac{1}{2}i\omega t - i\Omega t} \end{pmatrix} e^{-i\lambda_+ t} e^{i\tilde{q}z} + \begin{pmatrix} v_1^- e^{-\frac{1}{2}i\omega t - i\Omega t} \\ v_2^- e^{\frac{1}{2}i\omega t - i\Omega t} \end{pmatrix} e^{-i\lambda_- t} e^{iqz} \tag{12}$$

with $\lambda_+ = \lambda_- = -½ m\omega$. Such a choice determines the value of $q$ via Eq. (10).Then

$$\begin{pmatrix} \Psi^\uparrow \\ \Psi^\downarrow \end{pmatrix} = \sum_m e^{-i\Omega t} \begin{pmatrix} e^{i\frac{m-1}{2}\omega t} (K_-(c_1^m e^{iq_m z} + c_2^m e^{-iq_m z}) + K_+(c_3^m e^{i\tilde{q}_m z} + c_4^m e^{-i\tilde{q}_m z})) \\ e^{i\frac{m+1}{2}\omega t} (c_1^m e^{iq_m z} + c_2^m e^{-iq_m z} + c_3^m e^{i\tilde{q}_m z} + c_4^m e^{-i\tilde{q}_m z}) \end{pmatrix} \tag{13}$$

where we set the following values of $q_m,\ \tilde{q}_m$:

$$\lambda_- = \varepsilon_{xy} - \Omega + \tfrac{q_m^2}{2} - \tfrac{1}{2}\sqrt{(1-\omega)^2+4J_1^2} = -\tfrac{1}{2}\omega m \quad \Rightarrow$$
$$q_m^2 = 2\Omega - 2\varepsilon_{xy} + \sqrt{(1-\omega)^2+4J_1^2} - \omega m,\ m \le \frac{2\Omega - 2\varepsilon_{xy} + \sqrt{(1-\omega)^2+4J_1^2}}{\omega} \tag{14}$$

$$\lambda_+ = \varepsilon_{xy} - \Omega + \tfrac{\tilde{q}_m^2}{2} + \tfrac{1}{2}\sqrt{(1-\omega)^2+4J_1^2} = -\tfrac{1}{2}\omega m \quad \Rightarrow$$
$$\tilde{q}_m^2 = 2\Omega - 2\varepsilon_{xy} - \sqrt{(1-\omega)^2+4J_1^2} - \omega m,\ m \le \frac{2\Omega - 2\varepsilon_{xy} - \sqrt{(1-\omega)^2+4J_1^2}}{\omega} \tag{15}$$

The inequalities provide for real values of $q_m, \tilde{q}_m$.

Similarly, we solve the system (6) for $0 \le z \le b$.

$$\begin{pmatrix} \Phi^\uparrow \\ \Phi^\downarrow \end{pmatrix} = \sum_m e^{-i\Omega t + im\omega t/2} \begin{pmatrix} L_-(g_1^m e^{ip_m z} + g_2^m e^{-ip_m z}) + L_+(g_3^m e^{i\tilde{p}_m z} + g_4^m e^{-i\tilde{p}_m z}) \\ g_1^m e^{ip_m z} + g_2^m e^{-ip_m z} + g_3^m e^{i\tilde{p}_m z} + g_4^m e^{-i\tilde{p}_m z} \end{pmatrix}. \tag{16}$$

In this expression, we use the following notation:

$$L_{\pm} = \frac{i\beta k_y}{\alpha k_x \pm \sqrt{\alpha^2 k_x^2 + \beta k_y^2}}$$

$$p_m^2 = 2\Omega + 2U_0 + \sqrt{\alpha^2 k_x^2 + \beta k_y^2} - \omega m,\ m \le \frac{2\Omega + 2U_0 + \sqrt{\alpha^2 k_x^2 + \beta k_y^2}}{\omega} \quad (17)$$

$$\tilde{p}_m^2 = 2\Omega + 2U_0 - \sqrt{\alpha^2 k_x^2 + \beta k_y^2} - \omega m,\ m \le \frac{2\Omega + 2U_0 - \sqrt{\alpha^2 k_x^2 + \beta k_y^2}}{\omega}$$

Like in (14),(15), the inequalities in (17) provide for real values of $p_m, \tilde{p}_m$.

We will demonstrate that we need only a finite number of terms in (13) and (16) to satisfy matching at the interface and boundary conditions.

As a result, we get the following solution of the systems (5, 6):

$$\begin{pmatrix} \Psi^{\uparrow} \\ \Psi^{\downarrow} \end{pmatrix} = \sum_{j=0}^{n} e^{-i\Omega t} \times$$

$$\times \begin{pmatrix} e^{i\frac{m+2j-1}{2}\omega t} (K_-(c_1^{m+2j} e^{iq_{m+2j}z} + c_2^{m+2j} e^{-iq_{m+2j}z}) + K_+(c_3^{m+2j} e^{i\tilde{q}_{m+2j}z} + c_4^{m+2j} e^{-i\tilde{q}_{m+2j}z})) \\ e^{i\frac{m+2j+1}{2}\omega t} (c_1^{m+2j} e^{iq_{m+2j}z} + c_2^{m+2j} e^{-iq_{m+2j}z} + c_3^{m+2j} e^{i\tilde{q}_{m+2j}z} + c_4^{m+2j} e^{-i\tilde{q}_{m+2j}z}) \end{pmatrix} \quad (18)$$

$$\begin{pmatrix} \Phi^{\uparrow} \\ \Phi^{\downarrow} \end{pmatrix} = \sum_{j=0}^{n-1} e^{-i\Omega t + i\frac{(m+1+2j)\omega t}{2}} \times$$

$$\times \begin{pmatrix} L_-(g_1^{m+1+2j} e^{ip_{m+1+2j}z} + g_2^{m+1+2j} e^{-ip_{m+1+2j}z}) + L_+(g_3^{m+1+2j} e^{i\tilde{p}_{m+1+2j}z} + g_4^{m+1+2j} e^{-i\tilde{p}_{m+1+2j}z}) \\ g_1^{m+1+2j} e^{ip_{m+1+2j}z} + g_2^{m+1+2j} e^{-ip_{m+1+2j}z} + g_3^{m+1+2j} e^{i\tilde{p}_{m+1+2j}z} + g_4^{m+1+2j} e^{-i\tilde{p}_{m+1+2j}z} \end{pmatrix} \quad (19)$$

The number of terms in (19) is always one less than the number of terms in (18). We named this solution as a 'solution of the length *n*'. Matching and boundary conditions give the equations for the coefficients $c_l^k, g_l^k$. The continuity of the solution and its' first derivatives at $z = 0$ for arbitrary $j$, $j$=0, 1, …, $n$ lead to the equations:

$$\begin{aligned}
& c_1^{m+2j} + c_2^{m+2j} + c_3^{m+2j} + c_4^{m+2j} = g_1^{m+2j+1} + g_2^{m+2j+1} + g_3^{m+2j+1} + g_4^{m+2j+1} \\
& iq_{m+2j}c_1^{m+2j} - iq_{m+2j}c_2^{m+2j} + i\tilde{q}_{m+2j}c_3^{m+2j} - i\tilde{q}_{m+2j}c_4^{m+2j} = \\
& \quad = ip_{m+2j+1}g_1^{m+2j+1} - ip_{m+2j+1}g_2^{m+2j+1} + i\tilde{p}_{m+2j+1}g_3^{m+2j+1} - i\tilde{p}_{m+2j+1}g_4^{m+2j+1} \\
& K^-c_1^{m+2j+2} + K^-c_2^{m+2j+2} + K^+c_3^{m+2j+2} + K^+c_4^{m+2j+2} = L^-g_1^{m+2j+1} + L^-g_2^{m+2j+1} + L^+g_3^{m+2j+1} + L^+g_4^{m+2j+1} \\
& iq_{m+2j+2}K^-c_1^{m+2j+2} - iq_{m+2j+2}K^-c_2^{m+2j+2} + i\tilde{q}_{m+2j+2}K^+c_3^{m+2j+2} - i\tilde{q}_{m+2j+2}K^+c_4^{m+2j+2} = \\
& \quad = ip_{m+2j+1}L^-g_1^{m+2j+1} - ip_{m+2j+1}L^-g_2^{m+2j+1} + i\tilde{p}_{m+2j+1}L^+g_3^{m+2j+1} - i\tilde{p}_{m+2j+1}L^+g_4^{m+2j+1}
\end{aligned} \quad (20)$$

The boundary conditions at $z = b$ are following:

$$g_1^{m+2j+1}e^{ip_{m+2j+1}b}+g_2^{m+2j+1}e^{-ip_{m+2j+1}b}+g_3^{m+2j+1}e^{i\tilde{p}_{m+2j+1}b}+g_4^{m+2j+1}e^{-i\tilde{p}_{m+2j+1}b}=0$$
$$g_1^{m+2j+1}L^-e^{ip_{m+2j+1}b}+g_2^{m+2j+1}L^-e^{-ip_{m+2j+1}b}+g_3^{m+2j+1}L^+e^{i\tilde{p}_{m+2j+1}b}+g_4^{m+2j+1}L^+e^{-i\tilde{p}_{m+2j+1}b}=0 \quad (21)$$

Equations (20) and (21) relate $g_l^{m+2j+1}$ with $c_l^{m+2j}$ and $c_l^{m+2j+2}$, $l=1,2,3,4$. These equations can be represented in matrix form as follows:

$$\begin{pmatrix} 1 & 1 & 1 & 1 \\ iq_{m+2j} & -iq_{m+2j} & i\tilde{q}_{m+2j} & -i\tilde{q}_{m+2j} \\ 0 & 0 & 0 & 0 \\ 0 & 0 & 0 & 0 \end{pmatrix}\begin{pmatrix} c_1^{m+2j} \\ c_2^{m+2j} \\ c_3^{m+2j} \\ c_4^{m+2j} \end{pmatrix}=$$
$$=\begin{pmatrix} 1 & 1 & 1 & 1 \\ ip_{m+2j+1} & -ip_{m+2j+1} & i\tilde{p}_{m+2j+1} & -i\tilde{p}_{m+2j+1} \\ L^-e^{ip_{m+2j+1}b} & L^-e^{-ip_{m+2j+1}b} & L^+e^{i\tilde{p}_{m+2j+1}b} & L^+e^{-i\tilde{p}_{m+2j+1}b} \\ e^{ip_{m+2j+1}b} & e^{-ip_{m+2j+1}b} & e^{i\tilde{p}_{m+2j+1}b} & e^{-i\tilde{p}_{m+2j+1}b} \end{pmatrix}\begin{pmatrix} g_1^{m+2j+1} \\ g_2^{m+2j+1} \\ g_3^{m+2j+1} \\ g_4^{m+2j+1} \end{pmatrix} \quad (22)$$

$$\begin{pmatrix} K^- & K^- & K^+ & K^+ \\ iq_{m+2j+2}K^- & -iq_{m+2j+2}K^- & i\tilde{q}_{m+2j+2}K^+ & -i\tilde{q}_{m+2j+2}K^+ \\ 0 & 0 & 0 & 0 \\ 0 & 0 & 0 & 0 \end{pmatrix}\begin{pmatrix} c_1^{m+2j+2} \\ c_2^{m+2j+2} \\ c_3^{m+2j+2} \\ c_4^{m+2j+2} \end{pmatrix}=$$
$$=\begin{pmatrix} L^- & L^- & L^+ & L^+ \\ ip_{m+2j+1} & -ip_{m+2j+1} & i\tilde{p}_{m+2j+1} & -i\tilde{p}_{m+2j+1} \\ L^-e^{ip_{m+2j+1}b} & L^-e^{-ip_{m+2j+1}b} & L^+e^{i\tilde{p}_{m+2j+1}b} & L^+e^{-i\tilde{p}_{m+2j+1}b} \\ e^{ip_{m+2j+1}b} & e^{-ip_{m+2j+1}b} & e^{i\tilde{p}_{+2jm+1}b} & e^{-i\tilde{p}_{m+2j+1}b} \end{pmatrix}\begin{pmatrix} g_1^{m+2j+1} \\ g_2^{m+2j+1} \\ g_3^{m+2j+1} \\ g_4^{m+2j+1} \end{pmatrix} \quad (23)$$

The determinants of the matrixes in the right-hand side of (22) and (23) are

$$4(L^+-L^-)(\tilde{p}_{m+2j+1}\cos(b\tilde{p}_{m+2j+1})\sin(bp_{m+2j+1})-p_{m+2j+1}\cos(bp_{m+2j+1})\sin(b\tilde{p}_{m+2j+1})),$$
$$4L^+L^-(L^+-L^-)(\tilde{p}_{m+2j+1}\cos(b\tilde{p}_{m+2j+1})\sin(bp_{m+2j+1})-p_{m+2j+1}\cos(bp_{m+2j+1})\sin(b\tilde{p}_{m+2j+1})) \quad (24)$$

correspondingly. Therefore, the matrixes in the right-hand side (22) and (23) are non-degenerate with the exception of a countable set of the values of the coefficients of (3), (4). For all other values of the coefficients, we use these equations to obtain the coefficients of the solution (18), (19). Let's assume we get $c_l^{m+2j}, l=1,2,3,4$ at the preceding step. Then we obtain $g_l^{m+2j+1}, l=1,2,3,4$ from the equations (22). As soon as equations (22) are satisfied, the two last equations of (23) are satisfied as well, because the last two equations in (22) and (23) are identical. Choosing arbitrarily any pair of coefficients $c_l^{m+2j+2}, l=1,2,3,4$, for instance, $c_1^{m+2j+2}, c_2^{m+2j+2}$, we get the other two coefficients through the first two equations of (22). Then we repeat this procedure for the subsequent indices. In case the determinants of the matrices in the right-hand sides of (22), (23) equal zero, we chose one of $g_l^{m+2j+1}, l=1,2,3,4$ arbitrary. In that case the ranks of these matrices equal three and cannot be lesser, because in order to turn the all third minors of these matrices into zero, we should get $e^{2ibp_{m+2j+1}}=1$ and $e^{2ibp_{m+2j+1}}=-1$ simultaneously.

For the terms with minimal and maximal values of the indexes, we have additional conditions at the interface, $z = 0$. For $j=0$, we get

$$\begin{aligned} &K^-c_1^m + K^-c_2^m + K^+c_3^m + K^+c_4^m = 0 \\ &iq_m K^-c_1^m - iq_m K^-c_2^m + i\tilde{q}_m K^+c_3^m - i\tilde{q}_m K^+c_4^m = 0 \end{aligned} \qquad (25)$$

These two equations allow us to express, for instance, $c_3^m, c_4^m$ through $c_1^m, c_2^m$ . The only restriction is that we cannot set two coefficients to zero. Then we get $g_l^{m+1}, l=1,2,3,4$. For the maximal indexes, we have two additional equations as well. Therefore, we get four equations for $c_l^{m+2n}, l=1,2,3,4$, whereas for smaller $j$ we have only two equations.

$$\begin{aligned} &c_1^{m+2n} + c_2^{m+2n} + c_3^{m+2n} + c_4^{m+2n} = 0 \\ &iq_{m+2n}c_1^{m+2n} - iq_{m+2n}c_2^{m+2n} + i\tilde{q}_{m+2n}c_3^{m+2n} - i\tilde{q}_{m+2n}c_4^{m+2n} = 0 \\ &K^-c_1^{m+2n} + K^-c_2^{m+2n} + K^+c_3^{m+2n} + K^+c_4^{m+2n} = L^-g_1^{m+2n-1} + L^-g_2^{m+2n-1} + L^+g_3^{m+2n-1} + L^+g_4^{m+2n-1} \\ &iq_{m+2n}K^-c_1^{m+2n} - iq_{m+2n}K^-c_2^{m+2n} + i\tilde{q}_{m+2n}K^+c_3^{m+2n} - i\tilde{q}_{m+2n}K^+c_4^{m+2n} = \\ &= ip_{m+2n-1}L^-g_1^{m+2n-1} - ip_{m+2n-1}L^-g_2^{m+2n-1} + i\tilde{p}_{m+2n-1}L^+g_3^{m+2n-1} - i\tilde{p}_{m+2n-1}L^+g_4^{m+2n-1} \end{aligned} \qquad (26)$$

These equations give us $c_l^{m+2n}, l=1,2,3,4$. So, we have determined all coefficients. For $j=0,1,...,n-1$, we choose arbitrarily two out of four coefficients $c_l^{m+2j}, j=1,2,3,4$. These coefficients can be used to impose some additional conditions.

## IV. The structure of the solution

First, we note that upper indexes in the sum (18), (19) change by two. Solution of a length "*n*" can be represented as a linear combination of the solution of smallest length. The minimal length of a solution is 1. For this minimal solution, (19) consists of a single term, and (16) contains two terms. Therefore, our solutions can be divided into two groups, with odd and even upper indexes '*k*' of the coefficients $c_j^k, g_j^k$. A linear combination of the solutions belonging to one group belongs to the same group.

As an example, let us demonstrate the application of the foregoing scheme to a minimal solution. Let us now consider the following solution:

$$\begin{pmatrix} \Phi^{\uparrow} \\ \Phi^{\downarrow} \end{pmatrix} = \begin{pmatrix} e^{-i\Omega t + i\frac{(m+1)\omega t}{2}} (L_-(g_1^{m+1}e^{ip_{m+1}z} + g_2^{m+1}e^{-ip_{m+1}z}) + L_+(g_3^{m+1}e^{i\tilde{p}_{m+1}z} + g_4^{m+1}e^{-i\tilde{p}_{m+1}z})) \\ e^{-i\Omega t + i\frac{(m+1)\omega t}{2}} (g_1^{m+1}e^{ip_{m+1}z} + g_2^{m+1}e^{-ip_{m+1}z} + g_3^{m+1}e^{i\tilde{p}_{m+1}z} + g_4^{m+1}e^{-i\tilde{p}_{m+1}z}) \end{pmatrix} \qquad (27)$$

$$\begin{pmatrix}\Psi^{\uparrow}\\ \Psi^{\downarrow}\end{pmatrix}=e^{-i\Omega t}\begin{pmatrix} e^{i\frac{m-1}{2}\omega t}(K_-(c_1^m e^{iq_m z}+c_2^m e^{-iq_m z})+K_+(c_3^m e^{i\tilde{q}_m z}+c_4^m e^{-i\tilde{q}_m z}))\\ e^{i\frac{m+1}{2}\omega t}(c_1^m e^{iq_m z}+c_2^m e^{-iq_m z}+c_3^m e^{i\tilde{q}_m z}+c_4^m e^{-i\tilde{q}_m z})\end{pmatrix}+$$
$$+e^{-i\Omega t}\begin{pmatrix} e^{i\frac{m+1}{2}\omega t}(K_-(c_1^{m+2} e^{iq_{m+2} z}+c_2^{m+2} e^{-iq_{m+2} z})+K_+(c_3^{m+2} e^{i\tilde{q}_{m+2} z}+c_4^{m+2} e^{-i\tilde{q}_{m+2} z}))\\ e^{i\frac{m+3}{2}\omega t}(c_1^{m+2} e^{iq_{m+2} z}+c_2^{m+2} e^{-iq_{m+2} z}+c_3^{m+2} e^{i\tilde{q}_{m+2} z}+c_4^{m+2} e^{-i\tilde{q}_{m+2} z})\end{pmatrix} \quad (28)$$

The terms with time exponents $e^{-i\Omega t+i\frac{m-1}{2}\omega t}$ and $e^{-i\Omega t+i\frac{m+3}{2}\omega t}$ are present in (28) and are absent in (27). As a consequence, matching condition at the interface $z=0$ for these terms in (28) are following:

$$\begin{aligned}
&K_-(c_1^m+c_2^m)+K_+(c_3^m+c_4^m)=0\\
&iq_m K_-(c_1^m-c_2^m)+i\tilde{q}_m K_+(c_3^m-c_4^m)=0\\
&c_1^{m+2}+c_2^{m+2}+c_3^{m+2}+c_4^{m+2}=0\\
&iq_{m+2}(c_1^{m+2}-c_2^{m+2})+i\tilde{q}_{m+2}(c_3^{m+2}-c_4^{m+2})=0
\end{aligned} \quad (29)$$

The boundary conditions at $z=b$ is always zero:

$$\begin{aligned}
&L_-(g_1^{m+1}e^{ip_{m+1}b}+g_2^{m+1}e^{-ip_{m+1}b})+L_+(g_3^{m+1}e^{i\tilde{p}_{m+1}b}+g_4^{m+1}e^{-i\tilde{p}_{m+1}b})=0\\
&g_1^{m+1}e^{ip_{m+1}b}+g_2^{m+1}e^{-ip_{m+1}b}+g_3^{m+1}e^{i\tilde{p}_{m+1}b}+g_4^{m+1}e^{-i\tilde{p}_{m+1}b}=0
\end{aligned} \quad (30)$$

The matching of the other terms gives the following equations:

$$\begin{aligned}
&L_-(g_1^{m+1}+g_2^{m+1})+L_+(g_3^{m+1}+g_4^{m+1})=K_-(c_1^{m+2}+c_2^{m+2})+K_+(c_3^{m+2}+c_4^{m+2})\\
&p_{m+1}L_-(g_1^{m+1}-g_2^{m+1})+\tilde{p}_{m+1}L_+(g_3^{m+1}-g_4^{m+1})=q_{m+2}K_-(c_1^{m+2}-c_2^{m+2})+\tilde{q}_{m+2}K_+(c_3^{m+2}-c_4^{m+2})\\
&g_1^{m+1}+g_2^{m+1}+g_3^{m+1}+g_4^{m+1}=c_1^m+c_2^m+c_3^m+c_4^m\\
&p_{m+1}(g_1^{m+1}-g_2^{m+1})+\tilde{p}_{m+1}(g_3^{m+1}-g_4^{m+1})=q_m(c_1^m-c_2^m)+\tilde{q}_m(c_3^m-c_4^m)
\end{aligned} \quad (31)$$

Let us set some arbitrary values for $c_1^m, c_2^m$. Then the two first equations of (29) give

$$c_3^m=\frac{-K_-}{2K_+\tilde{q}_m}\left(c_1^m(\tilde{q}_m+q_m)+c_2^m(\tilde{q}_m-q_m)\right),\ c_4^m=\frac{-K_-}{2K_+\tilde{q}_m}\left(c_1^m(\tilde{q}_m-q_m)+c_2^m(\tilde{q}_m+q_m)\right) \quad (32)$$

Now, from the last two equations of (31) and equations (30) we get

$$\begin{aligned}
g_{1,2}^{m+1}&=\frac{e^{\mp ibp_{m+1}}(K_--K_+)\left(\mp i(c_1^m+c_2^m)\tilde{p}_{m+1}\cos(b\tilde{p}_{m+1})\pm(c_1^m-c_2^m)q_m\sin(b\tilde{p}_{m+1})\right)}{2K_+\left(\tilde{p}_{m+1}\cos(b\tilde{p}_{m+1})\sin(bp_{m+1})-p_{m+1}\cos(bp_{m+1})\sin(b\tilde{p}_{m+1})\right)}\\
g_{3,4}^{m+1}&=\frac{e^{\mp ib\tilde{p}_{m+1}}(K_--K_+)\left(\mp i(c_1^m+c_2^m)p_{m+1}\cos(bp_{m+1})\pm(c_1^m-c_2^m)q_m\sin(bp_{m+1})\right)}{2K_+\left(-\tilde{p}_{m+1}\cos(b\tilde{p}_{m+1})\sin(bp_{m+1})+p_{m+1}\cos(bp_{m+1})\sin(b\tilde{p}_{m+1})\right)}
\end{aligned} \quad (33)$$

The upper signs are related to $g_1^{m+1}, g_3^{m+1}$, whereas the lower signs are related to $g_2^{m+1}, g_4^{m+1}$.

Now we need to solve the two last equations of (29) and the first two equations of (31). We get

$$\begin{aligned}
c_1^{m+2} &= \frac{p_{m+1}L_-(g_1^{m+1}-g_2^{m+1})+\tilde{p}_{m+1}L_+(g_3^{m+1}-g_4^{m+1})+q_{m+2}(L_-(g_1^{m+1}+g_2^{m+1})+L_+(g_3^{m+1}+g_4^{m+1}))}{2(K_--K_+)q_{m+2}} \\
c_2^{m+2} &= \frac{\tilde{p}_{m+1}L_+(g_4^{m+1}-g_3^{m+1})+q_{m+2}L_+(g_3^{m+1}+g_4^{m+1})+L_-g_1^{m+1}(q_{m+2}-p_{m+1})+L_-g_2^{m+1}(p_{m+1}+q_{m+2})}{2(K_--K_+)q_{m+2}} \\
c_3^{m+2} &= -\frac{p_{m+1}L_-(g_1^{m+1}-g_2^{m+1})+\tilde{p}_{m+1}L_+(g_3^{m+1}-g_4^{m+1})+\tilde{q}_{m+2}(L_-(g_1^{m+1}+g_2^{m+1})+L_+(g_3^{m+1}+g_4^{m+1}))}{2(K_--K_+)\tilde{q}_{m+2}} \\
c_4^{m+2} &= \frac{\tilde{p}_{m+1}L_+(g_3^{m+1}-g_4^{m+1})+\tilde{q}_{m+2}L_+(g_3^{m+1}+g_4^{m+1})+L_-g_1^{m+1}(p_{m+1}-\tilde{q}_{m+2})-L_-g_2^{m+1}(p_{m+1}+\tilde{q}_{m+2})}{2(K_--K_+)\tilde{q}_{m+2}}
\end{aligned} \tag{34}$$

All we need in order to get final expressions is to substitute (33) into (34). Therefore, we express all coefficients in (27), (28) through two constants $c_1^m, c_2^m$ which can be chosen arbitrarily. Therefore, the system of the equations (29), (30), (31) is not overdetermined. This is a general situation for a solution of arbitrary length.

Let us correlate our solution with Floquet theory. Solution (18, 19) is a linear combination of the solutions corresponding to different values of $q_{m+2j}$ (13), (14), namely, $q_{m+2j}$ and $\tilde{q}_{m+2j}$ i.e., these are the solutions of different systems of ordinary differential equations. For a fixed $q_{m+2j}$, we have Floquet system of two equations. However, we use only one of two linearly independent solutions. The condition $\lambda_- = \lambda_+ = -\omega m/2$ implies that we use the solutions $q_m$ and $\tilde{q}_m$ for two different *m*.

It follows from the procedure of the construction of the solution that the number of constants which can be chosen arbitrarily for a solution of length *n* is 2*n*. Nevertheless, we cannot choose boundary conditions similar to those for a stationary problem for a multilayered system (see, e.g., Eq.(52) in [13]). Common boundary condition for a stationary problem, that is a multilayer with magnetic layers with fixed magnetization, are following. Only one $e^{+iqz}$ which is called 'wave coming from the left' is present. The other exponents are 'reflected', that is $e^{-iqz}$. The class of the solutions which we presented here inevitably contains more than one 'wave coming from the left' exponent. The equations for the coefficients do not admit a solution with only a single wave incident from the left.

## V. Conclusion

We constructed an exact solution to a system of two partial differential equations with two independent variables (*z*- and *t*-) and piecewise continuous coefficients in z- variable. The coefficients depend on time harmonically in one segment. This problem goes back to Schrödinger equation for two-layer magnetic system with rotating magnetization. In contrast with the common approach which uses infinite time series, we constructed an exact solution containing an arbitrary but finite number of time exponents. We analyzed the structure of the obtained solution and discussed valid boundary conditions.